# Leakage Suppression Across Temperature in $Al_{1-x}Sc_xN$ Thin Film Ferroelectric Capacitors through Boron Incorporation

Pedram Yousefian, Xiaolei Tong, Jonathan Tan, Dhiren K. Pradhan, Deep Jariwala, and Roy H. Olsson III

*Abstract*— This paper presents high-temperature ferroelectric characterization of 40 nm $Al_{1-x-y}B_xSc_yN$ (AlBScN) thin film capacitors grown by co-sputtering $Al_{0.89}B_{0.11}$ and Sc targets onto Pt(111)/Ti(002)/Si(100) substrates. Structural analysis confirmed a c-axis-oriented wurtzite structure with a low surface roughness of 1.37 nm. Ferroelectric switching, characterized by positive-up-negative-down (PUND) measurements up to 600 °C, exhibited a linear decrease in coercive fields from 6.2 MV/cm at room temperature to 4.2 MV/cm at 600 °C, while remanent polarization remained stable with temperature. Direct current I–V measurements highlight a significant suppression of leakage currents, over two orders of magnitude lower compared to AlScN capacitors fabricated under similar conditions. These results position AlBScN thin films as strong candidates for ferroelectric applications in extreme environments.

## I. Introduction

Ferroelectric materials integrated with CMOS-compatible semiconductor technologies have been extensively explored for non-volatile memory, logic, and microelectromechanical systems (MEMS) due to their switchable polarization behavior, scalability, and compatibility with existing fabrication processes [1]-[4]. Aluminum scandium nitride ($Al_{1-x}Sc_xN$, AlScN) particularly stands out, offering high remanent polarization (>100 μC/cm²), CMOS process compatibility, low-temperature deposition suitable for back-end-of-line integration, and stable ferroelectric switching at temperatures approaching 1000 °C [5]-[9]. However, significant challenges remain, especially at elevated temperatures, notably high leakage currents and limited margins between coercive ($E_c$) and breakdown fields ($E_{BD}$), impacting electrical endurance and reliability under extreme environments.

Recent advances demonstrate that incorporating boron into aluminum nitrides, $Al_{1-x}B_xN$ and $Al_{1-x-y}B_xSc_yN$ (AlBScN), enhances bandgap energy, significantly reduces leakage currents, and stabilizes polarization switching mechanisms [10]-[15]. Atomic-scale investigations attribute these improvements to boron-induced structural disorder and transient antipolar states during polarization reversal, lowering the switching energy barriers and potentially disrupting conductive leakage pathways [16]. Such modifications thus combine the superior ferroelectric properties of AlScN, high polarization and reduced coercive fields, with the enhanced bandgap and minimized leakage current seen in $Al_{1-x}B_xN$ [14], [15].

In this work, we demonstrate that boron incorporation in co-sputtered AlScN thin film capacitors noticeably enhances high-temperature ferroelectric stability and significantly reduces leakage currents. Temperature-dependent electrical characterization up to 600 °C confirms stable polarization switching, systematically reduced coercive fields, and over two orders of magnitude lower leakage currents compared to conventional AlScN devices fabricated under analogous thicknesses and Sc alloying. These findings position AlBScN as a highly promising ferroelectric material for demanding thermal and electrical environments.

## II. Device Fabrication And Characterization

A 40 nm AlBScN thin film was deposited onto a 6-inch Si(100) substrate with a 100 nm Pt(111)/10 nm Ti bottom electrode stack by 150 kHz pulsed DC reactive co-sputtering from 100 mm diameter $Al_{0.89}B_{0.11}$ and Sc targets using an Evatec CLUSTERLINE 200 II system. The $Al_{0.89}B_{0.11}$ and Sc target powers were set at 900 W and 655 W, respectively aiming to achieve a boron concentration near the center of the wurtzite phase stability range [17]. Deposition was carried out at 350 °C under a process gas flow of 40 sccm $N_2$ and 15 sccm Ar. Following the AlBScN deposition, a 100 nm Ni top electrode was sputtered using a Lesker PVD75 DC/RF Sputterer and patterned via a lift-off process. Bottom electrodes were exposed by selectively wet etching the AlBScN layer using a 45 mol% KOH solution.

Structural properties and crystallographic orientation were evaluated by θ–2θ and ω scans using a Rigaku SmartLab X-ray diffractometer (XRD). Surface morphology was examined by atomic force microscopy (AFM) using a Bruker Dimension Icon system.

Manuscript received X; revised X; accepted X. This work is supported by the Defense Advanced Research Project Agency under the COFFEE program. The deposition, patterning, and characterizations of $Al_{1-x-y}B_xSc_yN$ thin films were performed at the Singh Center for Nanotechnology, which is funded under the NSF National Nanotechnology Coordinated Infrastructure Program (NNCI-1542153).

P. Yousefian, X. Tong, J. Tan, D. K. Pradhan, D. Jariwala, and R. H. Olsson III are with the Department of Electrical and Systems Engineering, University of Pennsylvania, USA (e-mail: rolsson@seas.upenn.edu).

Electrical measurements were conducted with a Keithley 4200A-SCS semiconductor parameter analyzer. All tests were performed by biasing the bottom electrodes while sensing current and charge from the 100 μm diameter top circular electrodes. The breakdown field ($E_{BD}$) was determined under a 10 kHz triangular voltage waveform, using the Keithley 4200A-SCS for $V_{BD} < 40$ V and a Radiant Precision Premier II system for $V_{BD} > 40$ V, with the breakdown voltage defined at the preset current compliance threshold. Temperature-dependent measurements were conducted in a vacuum probe station (HP1000V-MPS, Instec Inc.) at ~$10^{-3}$ Torr, with a heated stage for thermal control.

## III. EXPERIMENTAL RESULTS

The structural and morphological properties of the AlBScN film are shown in Fig. 1. The θ–2θ XRD scan of AlBScN film reveals a pronounced (002) reflection of AlBScN along with substrate peaks (Fig. 1(a)), confirming c-axis-oriented wurtzite crystalline structure. The rocking curve of the AlBScN (002) peak exhibits a full width at half maximum (FWHM) of 4°, presented in Fig. 1(b), indicating the relatively low out-of-plane mosaicity. AFM analysis presents a smooth and uniform surface morphology, characterized by a measured root-mean-square (RMS) roughness of 1.37 nm, as shown in Fig. 1(c).

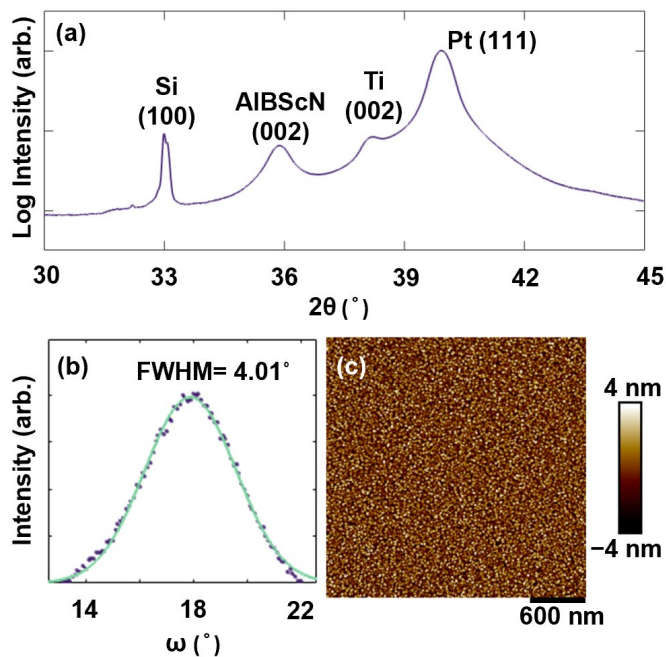

Fig. 1. (a) θ-2θ XRD scan of the 40 nm AlBScN thin film showing the (002) reflection, (b) Rocking curve of the (002) peak with a FWHM of 4°, (c) AFM topography of the AlBScN surface with a RMS roughness of 1.37 nm.

Ferroelectric switching characteristics were comprehensively evaluated using temperature-dependent J–E hysteresis loops and positive-up-negative-down (PUND) measurements. As shown in Fig. 2(a), J–E hysteresis loops measured at 10 kHz across temperatures from room temperature (RT) to 600 °C exhibit robust and clear ferroelectric switching throughout the measured temperature range. Coercive field ($E_c$) extracted from triangular PUND measurements at 10 kHz exhibit a linear decrease with increasing temperature, declining from 6.2 MV/cm at RT to 4.2 MV/cm at 600 °C under positive bias, and from -5.9 MV/cm to -3.2 MV/cm under negative bias, as summarized in Fig. 2(c). These temperature-dependent $E_c$ values align closely with prior literature findings [14], [15]. Additionally, Fig. 2(e) illustrates the stability of the ratio between the breakdown field ($E_{BD}$) and $E_c$, remaining consistent across the temperature range studied, thus ensuring a reliable operating margin even at elevated temperatures.

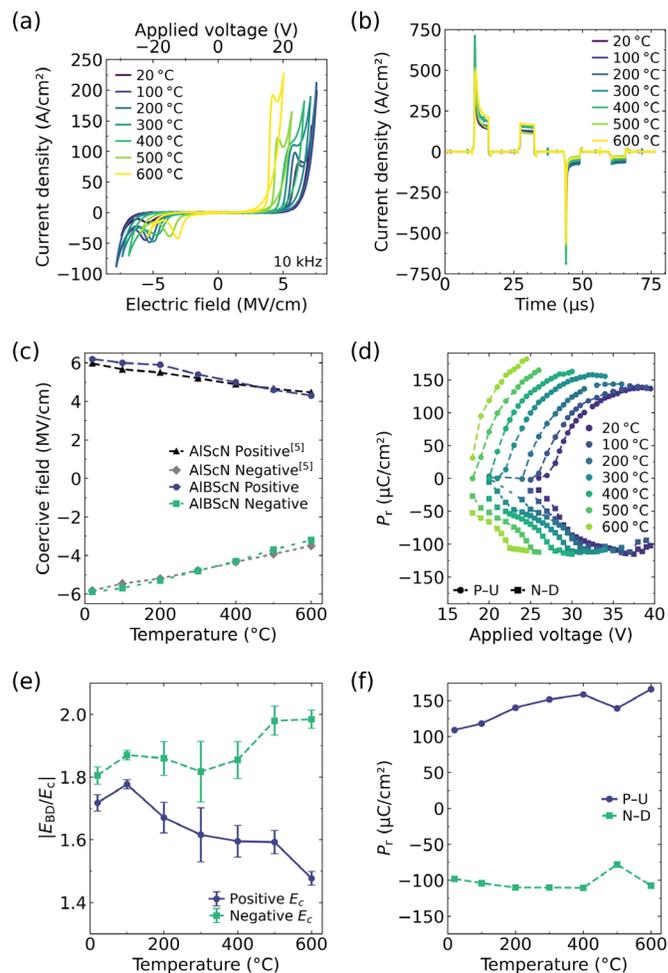

Fig. 2. (a) J–E hysteresis loops of AlBScN capacitors measured at 10 kHz across selected temperatures. (b) Current responses from rectangular PUND measurements at selected temperatures. (c) Comparison of positive and negative $E_c$ extracted from triangular PUND measurements as a function of temperature for the AlBScN capacitor and a previously reported AlScN capacitor [5], both of similar thickness and Sc alloying. (d) Voltage-dependent $P_r$ at different temperatures. (e) Ratio of $E_{BD}/E_c$ versus temperature. (f) Saturated $P_r$ extracted from 1/3 pulse width of rectangular PUND sequences for PU and ND pulses as a function of temperature.

Rectangular PUND measurements were performed to precisely determine remanent polarization ($P_r$). Each pulse sequence consisted of an 800 ns rise/fall time, a 5 μs pulse width, and an inter-pulse delay of 10 μs. Representative current density responses across selected temperatures are shown in Fig. 2(b). To ensure the reported $P_r$ values represent fully saturated polarization states, the voltage amplitude was progressively increased at each temperature until saturation was confirmed (Fig. 2(d)). The saturated $P_r$ values, obtained by integrating the subtracted current over one-third of the pulse width, are summarized in Fig. 2(f). The $P_r$ values from the positive-up (PU) pulses exhibited an increase from $109\,\mu C/cm^2$ at RT to $166\,\mu C/cm^2$ at $600\,°C$. This observed polarization value at $600\,°C$ is inconclusive because of the larger leakage currents observed at higher temperatures, which contributes to the polarization response [6]. In contrast, the negative-down (ND) sequence exhibited stable $P_r$ values, varying from $-98\,\mu C/cm^2$ at RT to $-107.5\,\mu C/cm^2$ at $600\,°C$.

Leakage current characteristics of the AlBScN capacitors were assessed through temperature-dependent direct current (DC) I–V measurements conducted up to $600\,°C$ (Fig. 3(a)). A comparative analysis in Fig. 3(b) highlights the substantial suppression of leakage current achieved with boron incorporation. AlBScN capacitors demonstrated more than two orders of magnitude lower leakage current compared to AlScN devices fabricated under similar Sc concentrations, thickness and processing conditions [5]. This notable reduction in leakage currents strongly suggests the effectiveness of boron doping in mitigating leakage conduction pathways while maintaining robust ferroelectric behavior.

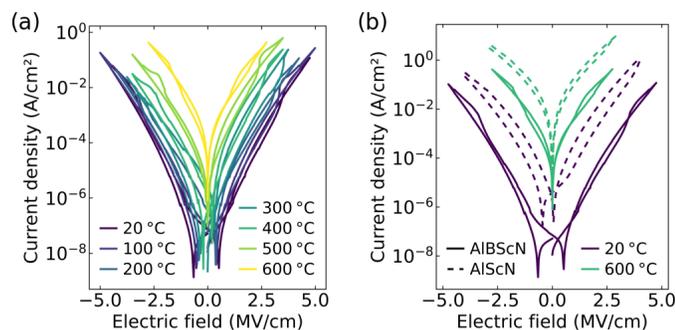

Fig. 3. (a) Temperature-dependent DC I–V curves of the AlBScN capacitor up to $600\,°C$. (b) Leakage current comparison between the AlBScN capacitor and a previously reported AlScN capacitor [5], both fabricated under similar conditions.

## IV. Conclusion

In this study, we successfully fabricated and characterized 40 nm AlBScN ferroelectric thin film capacitors grown via co-sputtering from $Al_{0.89}B_{0.11}$ alloy and Sc targets onto Pt(111)/Ti(002)/Si(100) substrates. Structural characterization confirmed c-axis-oriented wurtzite growth with a (002) preferred orientation and a low surface roughness of 1.37 nm. Temperature-dependent ferroelectric measurements demonstrated stable switching behavior up to $600\,°C$, with $E_c$ decreasing linearly from $6.2\,MV/cm$ to $4.2\,MV/cm$ while maintaining a consistent $E_{BD}/E_c$ ratio. Remanent polarization values in the ND sequence remained relatively constant, varying from $-98\,\mu C/cm^2$ at RT to $-107.5\,\mu C/cm^2$ at $600\,°C$ owing to the low leakage achieved via boron incorporation. Direct current I–V measurements revealed that boron incorporation reduced leakage currents by more than two orders of magnitude compared to AlScN capacitors fabricated at similar thickness and Sc alloying, while the $E_c$ of the two materials was nearly identical. These results establish AlBScN as a promising ferroelectric material for high-temperature and high-field applications, combining high polarization, reduced switching fields, and significantly enhanced leakage performance. Further optimization of the boron concentration and processing conditions may enable even greater stability and reliability of these capacitors for extreme operating environments.